\documentclass[review]{elsarticle}

\usepackage{amsmath,amssymb,amsfonts}
\usepackage{algorithmic}
\usepackage{float}
\usepackage{siunitx}

\usepackage{graphicx}
\usepackage{caption}
\usepackage{subcaption}
\usepackage[usenames]{color}
\definecolor{blue}{RGB}{0,0,255}

\journal{Renewable Energy}

\begin{document}
	
\begin{frontmatter}

\title{Power feedback strategy based on efficiency trajectory analysis for HCPV sun tracking}

\author[1]{Manuel G. Satu\'e}
\ead{mgarrido16@us.es}
\address{Escuela T\'ecnica Superior de Ingenier\'ia, Universidad de Sevilla, Sevilla, Spain}
\author[2]{Fernando Casta\~no}
\ead{fercas@us.es}
\address{Escuela T\'ecnica Superior de Ingenier\'ia, Universidad de Sevilla, Sevilla, Spain}
\author[3]{Manuel G. Ortega}
\ead{mortega@us.es}
\address{Escuela T\'ecnica Superior de Ingenier\'ia, Universidad de Sevilla, Sevilla, Spain}
\author[4]{Francisco R. Rubio}
\ead{rubio@us.es}
\address{Escuela T\'ecnica Superior de Ingenier\'ia, Universidad de Sevilla, Sevilla, Spain}

\begin{abstract}

This paper presents a control strategy for sun trackers which adapts continuously to different sources of error, avoiding the necessity of any kind of calibration by  analyzing the produced electric power to sense the position of the Sun. The proposed strategy is able to meet the strict specifications for HCPV sun trackers despite of mechanical uncertainties (misalignments in the structure itself, misalignment of the solar modules with respect to the wing, etc.) and installation uncertainties (misalignments of the platform with respect to geographical north). Experimental results with an industrial-grade solar tracker showing the validity of the proposed control strategy under sunny and moderate cloudy conditions, as well as with different installation precisions by un-calibrating the system on purpose are exposed.

\end{abstract}


\end{frontmatter}

%\linenumbers

\section {Introduction} \label{sec:Intro}

By using high concentration photovoltaic modules (HCPV), the efficiency of the transformation of light energy into electrical energy can increase twice as much as the provided by regular photovoltaic modules \cite{ZUBI20092645}. HCPV modules can be composed of two king of solar cells: high efficiency silicon and multijunction photovoltaic cells. The use of lenses which concentrate the sun rays on to multi-junction solar cells is what allows for increased efficiency \cite{PEREZHIGUERAS20111810}.

\begin{figure}
	\centering
	\begin{subfigure}[b]{0.45\textwidth}
		\includegraphics[width=\textwidth]{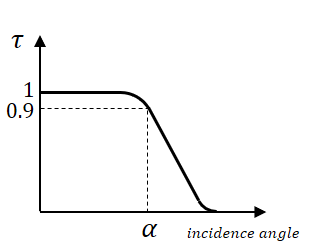}
		\caption{}
		\label{fig:ModeloLente_a}
	\end{subfigure}
	~ \quad  %add desired spacing between images, e. g. ~, \quad, \qquad, \hfill etc. 
	\begin{subfigure}[b]{0.4\textwidth}
		\includegraphics[width=\textwidth]{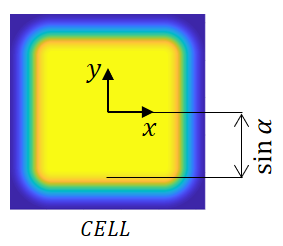}
		\caption{}
		\label{fig:ModeloLente_b}
	\end{subfigure}
	\caption{Concentrator model. \ref{fig:ModeloLente_a}) Definition of half-acceptance angle. \ref{fig:ModeloLente_b}) Efficiency values on the $x-y$ plane of a virtual cell.}
	\label{fig:ModeloLente}	
\end{figure}

HCPV modules concentrate the sun rays on to the solar cells using Fresnel lenses, which are characterized by their half-acceptance angle, $\alpha$. The half-acceptance angle is defined as the maximum angle at which incoming sun rays still fall upon the photovoltaic cells so that and the photoelectric effect occurs. Due to fabrication imperfections, in practice, the half-acceptance angle is defined as the angle for which efficiency drops to 90\% of its maximum, as shown if Fig. \ref{fig:ModeloLente_a}.

The half-acceptance angle of HCPV panels is typically around 1 degree \cite {YAVRIAN}, being this fact a very important restriction for the design  of the HCPV sun trackers and its control system. The system must have a very high pointing precision in order to keep the solar modules efficiency around its maximum and thus not waste energy. To achieve the high aiming precision requirement, HCPV sun trackers use multiple gearboxes and encoders of thousand of pulses per turn, but there are some uncertainties that decrease the pointing accuracy. This uncertainties do not depend on the mechanical design or the different components that constitute the solar tracker, but of its installation, its assembly, etc. The sources of uncertainty that can be found in a sun tracking system can be divided into two categories: hardware and software. Regarding the hardware, there are errors related to foundation and mechanical assembly. Misalignment of the sun tracker azimuth zero with respect to geographical north and misalignment of the elevation zero with respect the skyline are caused by a not too precise foundation of the sun tracker. Misalignment of the modules with respect to the wing is caused by a not too accurate assembly or by aging of the plant. This last kind of misalignment causes electrical mismatches that distorts the I-V output of the modules array, decreasing the efficiency of the sun tracker \cite{RODRIGO2016374}. Concerning software, there can be errors related to a bad adjustment of the zero of the encoders, a bad calibration of the system or even lack of calibration. Also included in this category are the errors introduced by the Solar Equations, such as the geographical positioning error of the sun tracker, the clock error of the controller unit and the accuracy of the algorithm itself.

There are two strategies to track the Sun: open-loop by means of Solar Equations and closed-loop, which uses some kind of sensor to locate the Sun. Some examples of open-loop controlled sun trackers are  \cite{ABDALLAH20041931} and \cite{FATHABADI2016485}. Regarding closed-loop controlled sun trackers, \cite{ROTH2004393} designed and constructed an electromechanical system to track the position of the Sun using a four-quadrant photo detector to sense the position of the Sun. \cite{YAO201488} presented a dual-axis solar tracker with two operation modes, one for PV systems and another for HCPV systems, was built. The HCPV mode used a combination of open-loop and closed-loop controls. The closed-loop is meant for fine pointing, while the open-loop is used in cloudy conditions. The sensor used is a four-quadrant photocell circuit. In \cite{Seung2015} high precision solar tracking was achieved also by a combination of open-loop and closed-loop controls. It operates in open-loop by means of Solar Equations until the tracking error exceeds a threshold, and in that time instant it switches to closed-loop control. The feedback sensor consists of a camera that senses the position of the Sun by looking directly to it. \cite{FATHABADI201667} built and compared two sun trackers, one open-loop and one closed-loop. In order to close the loop an irradiance sensor equipped with a radiance limiting tube is mounted in a secondary dual-axis machanical system. \cite{GARRIDO2016689} propose a cascade closed-loop control, where the inner loop employs non-linear proportional - proportional integral (NP-PI) controller and the outer loop used a proportional integral (PI) controller, in order to improve the tracking accuracy and reduce the actuator wear. \cite{ABDOLLAHPOUR2018136} also used a camera to locate the Sun, but in this case the shadow projected over a plate by a shadow-casting object is analyzed to infer the position of the Sun. An interesting work by \cite{CARBALLO20191158} uses computer vision and deep learning for solar tracking. This approach can provide additional information for the sun tracking system control like cloud movements prediction, atmospheric attenuation or measures of concentrated solar radiation, which can improve the control strategies of the system and therefore the system efficiency. The control system was tested with a heliostat in CESA central tower system, located in Plataforma Solar de Almer\'ia, but the authors claims that the same approach can be used for other kind of sun trackers, as HCPV sun trackers.

As open-loop control strategies rely exclusively on Solar Equations to determine the position of the Sun,  this strategies are greatly affected by installation errors, errors introduced by the possible lack of accuracy of the Solar Equations and errors due to the apparent position of the Sun for certain conditions of the atmosphere caused by refraction. Closed-loop control strategies use electro-optical sensors as \cite{SOLARMEMS}, pyrheliometers, or other sensors that can provide the position of the Sun with respect to the reference frame of the sensor itself. Closed-loop control strategies are not affected by installation errors but by a bad assembly of the sensor on to the wing of the sun tracker, which can produce a misalignment between the reference frames of the wing (or solar modules) and the optical sensor. Generally, closed-loop strategies are best suited for HCPV sun tracking because they deliver smaller tracking errors, but the great disadvantage is that they do not perform well on cloudy days, as their input is a measure of the position of the Sun. In order to mitigate the sources of uncertainty related to the installation the common actions are to perform a precise installation of the sun tracker by specialized personnel, which is time consuming and costly, or to perform a system calibration after installation \cite{SATUE2020311}, which is a complex task.

This paper presents the design, implementation and experimental testing of a control algorithm for HCPV sun trackers which allows the auto-correction of the different error sources that affect the tracking system by analyzing the electrical power produced to provide a sun position measurement to close the control loop. To use explicitly the electric power produced as a sensor is important because  the effect of the refraction of solar radiation in the atmosphere can modify the apparent position of the Sun for certain conditions of the atmosphere \cite{Jenkins_2013}. On the other hand, using the variable to be maximized as an indirect measure of the position of the Sun is an advantage over other solar sensors, which may be misaligned with respect to the solar modules. Therefore, the need to perform an accurate and expensive installation of the equipment or the need to perform a initial or periodic calibrations in order to estimate the transform relations between different reference frames of the sun tracker is eliminated. The proposed control strategy is valid as far as the sun tracker is able to perform movements in the azimuth and  elevation coordinates in an independent manner. Although there are different possibilities for the kinematic configuration of a sun tracker, almost all commercial sun trackers have this two-axis kinematic configuration \cite{NGOPV2012}. The idea is to use the photovoltaic modules as a sensor to estimate the position of the Sun, and the  Solar Equations to predict the future position of the sun with the purpose of maintaining the pointing vector of the sun tracker ahead of the position of the Sun. By using the produced electric power as a measure of the position of the Sun, the control strategy is explicitly trying to maximize the collected energy. In addition, the algorithm allows to reduce the number of movements that the sun tracker performs throughout the day. This is achieved by maximizing the path of the solar beam projection on the plane of the solar cell, so that all the space within the limits of the maximum efficiency zone is used (see Fig. \ref{fig:ModeloLente_a}). To do this, it is necessary to predict the position of the Sun in the future and to use a simple model of the lens of the solar collector together in conjunction with the kinematic model of the solar tracker. Until the Sun projects outside the region of maximum efficiency, there will not be a new pointing movement.

The tracking of the position of the Sun is carried out by means of a closed-loop control strategy in which the Solar Equations are a feed-forward which provide approximate coordinates for the pointing of the wing of the sun tracker, while a controller applies a correction on these coordinates by feed-backing the instantaneous DC power produced to estimate the real position of the Sun.

This work is inspired by the algorithm proposed in \cite{RUBIO20072174}. The main difference of the algorithm presented in this paper with respect \cite{RUBIO20072174} is that the produced power trajectories are analyzed and certain estimation rules are applied in order to estimate the position of the sun instead of just searching for a maximum in the produced power, which is prone to errors due to noisy signals and other effects. On the other hand, the sun tracker used in \cite{RUBIO20072174} did not have HCPV solar modules, so the authors emulated its behavior by using conventional modules with tubes perpendicular to the modules cells. Furthermore, a constant not optimal electric load was used instead of a power inverter. Moreover, the used solar tracker was a low-cost domestic tracker. In this work, test results performed with a high concentration solar tracker of larger dimensions and nominal output electric power, such as those found in solar plants, are provided.

The main innovation provided by this work is to make possible the start-up of a industrial-grade HCPV solar tracker without the need for a fine adjustment through calibrations, and to keep it continuously correcting, by analyzing the electrical power produced to provide a sun position measurement to close the control loop.

The remainder of the paper is structured as follows: the sun tracker characteristics are described in Section \ref{sec:SunTrackerDescription}. Section \ref{sec:TrackingStrategy} explains the proposed control strategy to track the Sun. The devices used to build the control system and their relationships are described in Section \ref{sec:ControlSystem}. Experimental results, including a comparison with an open-loop strategy, are shown in Section \ref{sec:Results}. The main conclusions of this work are drawn in Section \ref{sec:Conclusiones}.

\section{Sun tracker specs} \label{sec:SunTrackerDescription}

\begin{figure}[h]
	\centering
	\includegraphics[scale = 0.4]{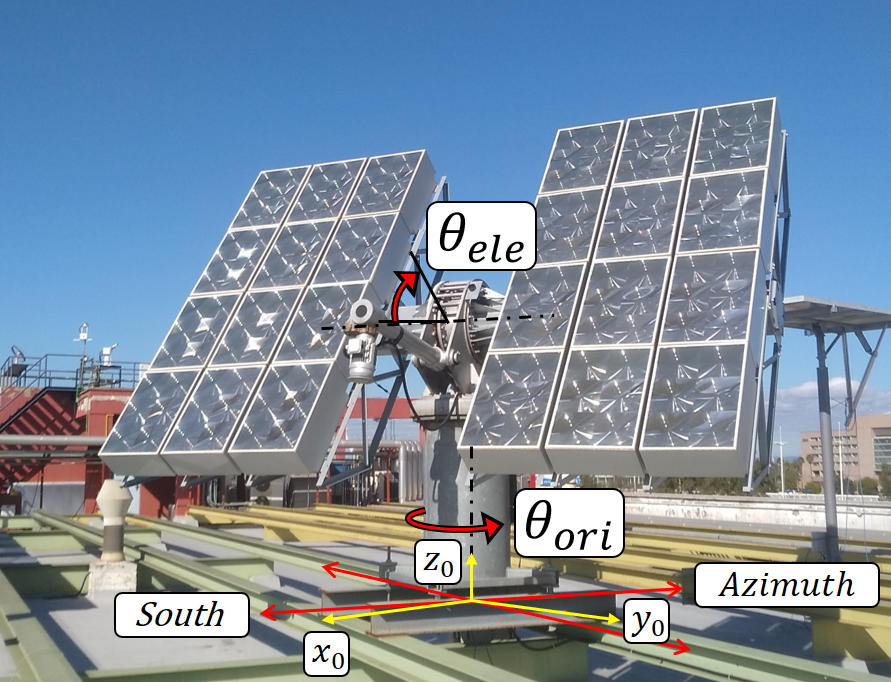}
	\caption{Sun Tracker used for the tests. Located on the roof of the Department of Systems and Automatic Control Engineering Laboratories at the University of Seville, Spain. $\theta_{ori}$ and $\theta_{ele}$ are sun tracker orientation and elevation coordinates, respectively. The reference system for the Solar Equations (red) and the reference system of the platform of the sun tracker (yellow) are not aligned.}
	\label{fig:SunTracker}	
\end{figure}

The sun tracker used in this work, depicted in Fig. \ref{fig:SunTracker}, is located on the roof of the Department of Systems and Automatic Control Engineering Laboratories at the University of Seville, Spain.

For the description of the installation that constitutes the solar tracker, it can be divided into the following parts: mechanical structure (fixed and mobile), generation and transformation devices, orientation movement control equipment and specific instrumentation equipment. 

The mechanical structure of the sun tracker has two degrees of freedom to follow the movement of the Sun in Azimuth and in elevation independently. It consists of  a fixed pole on which the azimuthal rotation mechanism (worm drive) is supported, which in turn supports the elevation mechanism (high accuracy linear actuator) and the wing. The HCPV modules are mounted on the wing.

To actuate on the spatial orientation of the wing there are two three-phase asynchronous motors (with a rated power of 550 W for the orientation motor and 750 W for the elevation motor) commanded by two variable frequency drives which are managed by a programmable logic controller (PLC). The orientation motor has two gearboxes attached in order to increase accuracy. The sun tracker orientation coordinate, $\theta_{ori}$, is measured by an encoder  with a resolution of $2^{14}$ pulses per revolution. The measure of the elevation coordinate, $\theta_{ele}$, is provided by an inclinometer with an accuracy of $\pm 0.1$ degrees. With this configuration, the sun tracker has an accuracy below 0.1 degrees in orientation movements and equal to 0.1 degrees for elevation movements.

The generator equipment is constituted by the HCPV modules and the transformation equipment is a power inverter. The installation has 24 solar modules with a total catchment area, $S_c$, of 9.3 $\mathrm{m^2}$. The electrical characteristics of the modules are shown in Table \ref{tab:ParamPanelISOFOTON}. The modules are connected in series, so that the voltage that can be reached in terminals will be approximately 443 V open circuit. The power inverter is a SMC Sunny Boy with a nominal power of 2500 W. Its input DC characteristics are a voltage between 260 and 500 volts and maximum amperage of 10 A. The output AC characteristics are 230 V (50 Hz) and 11 A.

\begin{table}[h!]	
	\begin{center}
		\begin{tabular}{|c|c|r|}
			\hline	
			Short-circuit current & $I_{sc}$ & 6.35 A \\
			Open-circuit voltage &  $V_{oc}$ & 18.45 V \\
			Power & $DC power$ & 95 W \\
			Max. power current & $I_{mp}$ & 5.73 A \\ 
			Max. power voltage & $V_{mp}$ & 16.62 V\\
			\hline	
			\multicolumn{3}{|c|}{ $83.6$ W  $3$ \%, $900 \; \mathrm{W/m^2}, \; T_c = 60 \; \si{\degree}$C} \\
			\hline	
		\end{tabular}
	\caption{Electrical characteristics of ISOFOTON GEN-2 solar module.}
\label{tab:ParamPanelISOFOTON}
\end{center}
\end{table}

As main control equipment, an Schneider MC80 programmable logic controller has been used in conjunction with a PC. The control algorithm determines the time instant at which the motors must rotate and generates the corresponding references for the variable frequency drives.

Regarding the specific instrumentation, a power meter to measure the instantaneous power generated by the installation on the direct current side and an optical sensor that provides the direct normal solar irradiance (dni) as well as the sun vector referred to the sensor reference system are installed.

\begin{figure}[h]
	\centering
	\includegraphics[scale = 0.5]{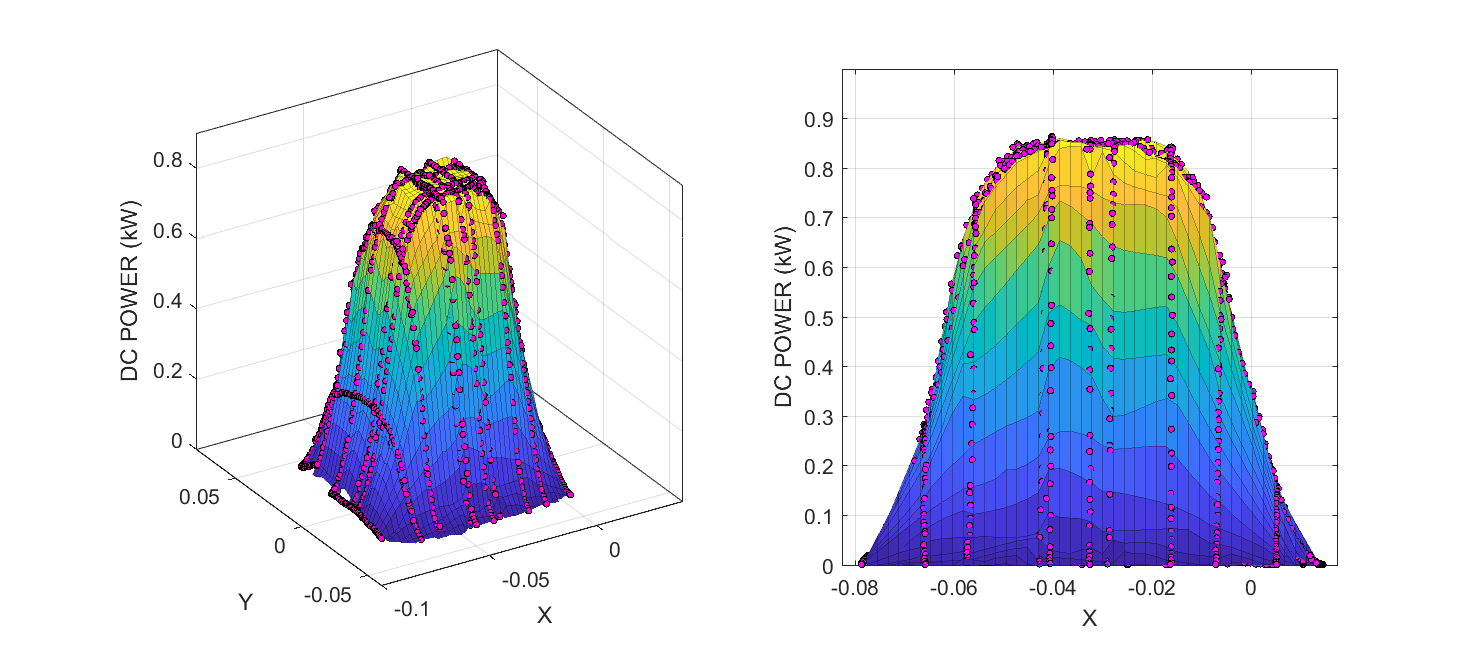}
	\caption{Power surface obtained scanning an area around the sun vector projection on the plane of a virtual cell. Sampled data is represented as magenta dots.}
	\label{fig:PowerSurface}	
\end{figure}

In order to measure the value of the semi-acceptance angle of the HCPV modules, an experiment was carried out. It consisted in closing the DC circuit with an electrical dissipation resistance (leaving the power inverter out of the circuit) and performing orientation and elevation movements while sampling the power produced by the HCPV modules (under almost constant irradiance) as well as the sun vector coordinates provided by the optical sensor. The result of the test is the power surface shown in Fig. \ref{fig:PowerSurface}. In this surface it is possible to measure the semi-acceptance angle, which is found to be approximately equal to 1.14 degrees for the ISOFOTON GEN-2 modules. This value is approximately an order of magnitude bigger than the pointing accuracy of the tracking system, which is a requirement for the proposed algorithm to work.

\section {Sun Tracking control strategy} \label{sec:TrackingStrategy}

\begin{figure}[h]
	\centering
	\includegraphics[scale = 0.3]{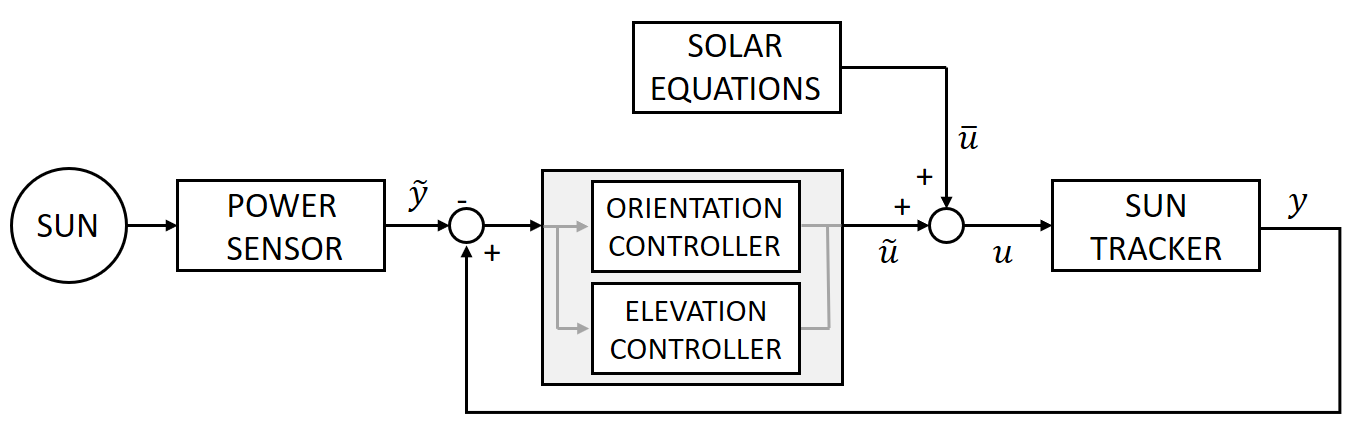}
	\caption{Control scheme.}
	\label{fig:DiagramaControl}	
\end{figure}

From a control theory point of view, the control strategy proposed in order to track the Sun consists on using the Solar Equations as a feed-forward that provides the approximate position of the Sun in a geographical coordinate system, $\bar{u}$, and closing the loop by using produced power measurements which will translate into sun position in the sun tracker coordinate system, $\tilde{y}$. A block diagram showing the control chain is depicted in figure \ref{fig:DiagramaControl}. The separate orientation and elevation controllers computes the correction, $\tilde{u}$, which will be applied to the the Solar Equations in order to generate orientation and elevation references for the sun tracker motors, $u$. These references are the inputs of two low level position controllers. This control scheme which computes orientation and elevation references is not executed in a continuous way, but in an asynchronous manner when a condition is fulfilled, as will be explained later.

The inputs of the Solar Equations are time, longitude and latitude. Its outputs are the Azimuth and elevation of the Sun. This work uses the PSA algorithm to estimate the position of the Sun \cite{BLANCOMURIEL2001431}. The need to use the closed-loop is due to the fact that there are errors in the estimation of the position of the Sun because of the causes explained in Section \ref{sec:Intro}.

As explained before, the produced electric power is used indirectly as a measure of the sun position. To achieve this goal, the sun tracker kinematic model and a simple model of a squared concentrator are used in conjunction with the corrected Solar Equations. These models are used to predict both the next reference for the sun tracker and the time instant on which the sun rays will exceed a certain angle, $\beta\:(\beta < \alpha)$, while the sun tracker is motionless at the previously calculated set-point. The main idea behind the previous prediction is that the sun tracker will move ahead of the sun and then will wait for the sun to travel the maximum possible distance over the photovoltaic cell in order to keep the sun beam inside the maximum power area of the virtual cell, which is defined by $\beta$. This strategy consisting in moving ahead the Sun also enables to minimize the number of movements that the sun tracker performs during a day. 

\begin{figure}
	\centering
	\begin{subfigure}[t]{0.45\textwidth}
		\includegraphics[width=\textwidth]{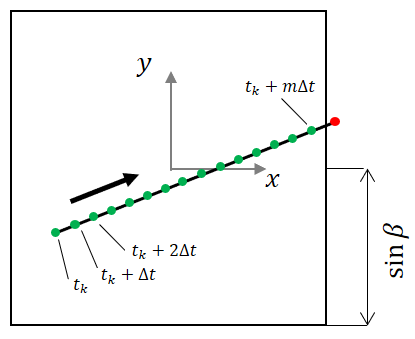}
		\caption{}
		\label{fig:ECS_Adelanto_a}
	\end{subfigure}
	~ \quad  %add desired spacing between images, e. g. ~, \quad, \qquad, \hfill etc. 
	\begin{subfigure}[t]{0.383\textwidth}
		\includegraphics[width=\textwidth]{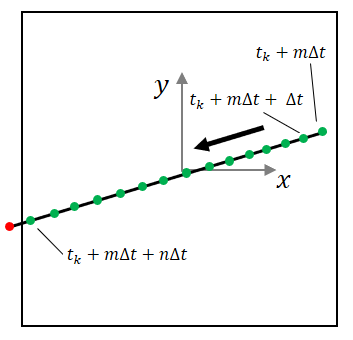}
		\caption{}
		\label{fig:ECS_Adelanto_b}
	\end{subfigure}
	\caption{Prediction using Solar Equations to point the HCPV modules ahead of the sun. \ref{fig:ECS_Adelanto_a}) Step one: motionless sun - mobile sun tracker.\ref{fig:ECS_Adelanto_b}) Step two: mobile sun - motionless sun tracker. Dots represents the projection of the sun vector on to the virtual solar cell.}
	\label{fig:ECS_Adelanto}
\end{figure}

The prediction consist of two steps. In the first one, using the models mentioned above, the spatial orientation of the wing of the sun tracker ahead of the position of the Sun still at current time and at the boundary of the region defined by $\beta$, is determined. To do so, first the sun position is calculated at the current time instant using the Solar Equations. The calculated sun position will not change during this first stage. Next, the time advances in a loop and the kinematic model of the sun tracker is used in conjunction with the concentrator model so that the wing points to the position given by Solar Equations evaluated at the advanced time. The loop breaks when the sun rays project outside the virtual squared cell as shown in Fig.  \ref{fig:ECS_Adelanto_a}. The time resolution in this loop is 1 second. The values of the sun tracker coordinates $(\theta_{ori},\theta_{ele})$ corresponding to the time instant just before said condition is fulfilled are stored as the next coordinates reference for the sun tracker, $(\theta_{ori},\theta_{ele})^{REF}$. In the second step of the prediction, the objective is to determine the time instant at which the sun rays projects outside the boundaries defined by $\beta$ while the simulated sun tracker stays motionless at the pose calculated in step one. The time keeps running in a loop while the sun position is actualized using the Solar Equations. As in the previous stage, the model of the concentrator is used to determine the time instant at which the sun rays project outside the virtual squared cell as depicted in Fig. \ref{fig:ECS_Adelanto_b}. The time instant just before the exiting condition is stored as the time of the next initiation of the control sequence, $t_{mov}$.

\begin{figure}[h]
	\centering
	\includegraphics[scale = 0.5]{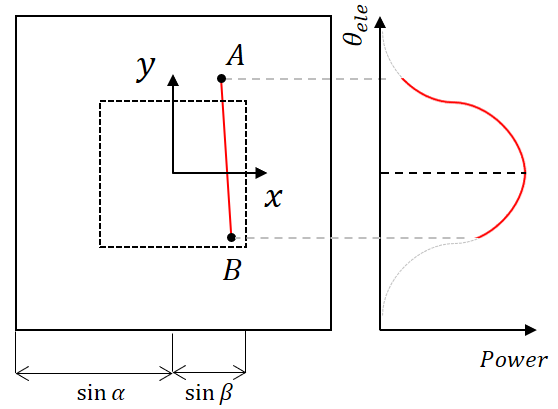}
	\caption{Idealized representation of the trajectory of the sun beam projected over the photovoltaic cell and its associated electric power trajectory during an elevation movement.}
	\label{fig:TrayectoriaSuperficiePotencia}	
\end{figure}

After this second step, the computed references $(\theta_{ori},\theta_{ele})^{REF}$ are given to the low level azimuth and elevation controllers in order to start a new pointing movement. The movement is not performed on both  axes at the same time.  First, the orientation movement is executed and, once it has finished, the elevation movement is executed. The following measures are sampled in separated tables for orientation and elevation while these movements are carried out:

\begin{itemize}
	\item produced DC electric power, $P$
	\item irradiance (DNI), $Irr$
	\item sun tracker coordinate, $\theta_{ori}$ or $\theta_{ele}$, depending on the type of movement
	\item time-stamp, $ts$
\end{itemize}

Figure \ref{fig:TrayectoriaSuperficiePotencia} shows an representation of the trajectory of the sun rays projected over the photovoltaic cell when the sun tracker performs an elevation movement. The trajectory starts at point $A$ and ends at point $B$, and during the movement of the sun tracker from $A$ to $B$, the electric power describes the curve shown. Although the movement only involves the coordinate $\theta_{ele}$, the trajectory may not be perfectly parallel to $y$ axis of the cell due to misalignment in the assembly of the solar modules to the wing, or installation error of the platform on the ground, or both.

Once the movements are finished, the controller processes the trajectories of the sampled variables using a non-causal finite impulse response (FIR) filter to reduce the noise \cite{GUSTAFSSON}, \cite{CETIN}. Then, the  instantaneous efficiency trajectories, $[ts,\eta]$, of both orientation and elevation movements are computed. The efficiency is determined with expression (\ref{eq:efficiency}).

\begin{equation}
\label{eq:efficiency}
\eta = \frac{P}{Irr \cdot S_c}
\end{equation}

The efficiency trajectories and sun tracker coordinate trajectories, $[ts,\theta]$, are analyzed with the purpose of estimating which was the position of the Sun (in sun tracker coordinates) while the sequence of movements was being executed, as well as an associated time-stamp, $\hat{ts}$, of the time instant in which the sun tracker had these coordinates. The use of the efficiency instead of the power for the analyzed trajectories is because the efficiency adds a filtering effect that smooths the shapes and mitigates the effect of the clouds on the shape of the trajectory. The estimation of the sun position in the analysis stage will be studied later in Section \ref{sec:TrajectoryAnalysis}. The time-stamp $\hat{ts}$ allows to evaluate the Solar Equations  so that the controller can compute the errors in azimuth and elevation that existed between the Solar Equations and the measurements in that past moment. These errors are the inputs of two asynchronous proportional and integral (PI) controllers that will try to reduce the discrepancies between the Solar Equations feed-forward and the estimated position from the measures. The output of the PI controllers are the offsets that will correct the Solar Equations the next time the controller decides to perform a sun pointing movement, which will happen when the controller clock reaches the time $t_{mov}$.

In the analysis stage, the trajectories associated to the azimuth and/or elevation movements may be discarded due either because they do not have enough samples (due to a very small associated movement), or because they are not classified within the expected types of trajectory (due to a cloudy period, or even because the power inverter is off). If this is the case, the associated PI controller will not calculate a new offset, but the last computed offset is used to correct the Solar Equations.

When both trajectories associated to the azimuth and elevation movements are accepted in the analysis stage, the time stamp $\hat{ts}$ is computed with expression (\ref{eq:time_stamp_medio}). If only one of the trajectories is accepted it is determined with  (\ref{eq:time_stamp_no_medio}).

\begin{equation}
\label{eq:time_stamp_medio}
\hat{ts} = t(\hat{\theta}_{ori}) + \frac{t(\hat{\theta}_{ele}) - t(\hat{\theta}_{ori})}{2}
\end{equation}

\begin{equation}
\label{eq:time_stamp_no_medio}
\hat{ts} = t(\hat{\theta})
\end{equation}

The PI controllers are discrete, work in an asynchronous manner and do not take into account the time interval elapsed between calls to them. Their associated equations are the following:

\begin{equation}
\label{eq:PI1a}
e_k^{azi} = \Phi_{azi}^{ST} - \Phi_{azi}^{SE}
\end{equation}

\begin{equation}
\label{eq:PI1b}
e_k^{ele} = \hat{\theta}_{ele} - \Phi_{ele}^{SE}
\end{equation}

\begin{equation}
\label{eq:PI2}
\tilde{u}_{k+1} = K_p \; e_k + K_i \sum_{k=0} e_{k-1}
\end{equation}

The angles $\Phi_{azi}^{SE}$ and $\Phi_{ori}^{SE}$ are obtained by evaluating the Solar Equations at time instant $\hat{ts}$.
The angle corresponding to the sun tracker coordinate $\hat{\theta}_{ori}$ is transformed to an azimuthal angle coordinate, $\Phi_{azi}^{ST}$, in order to be comparable with the azimuthal angle provided by Solar Equations, $\Phi_{azi}^{SE}$. 

\subsection{Efficiency trajectory analysis} \label{sec:TrajectoryAnalysis}

The goal of the trajectory analysis is to classify the efficiency trajectories sampled during the orientation and elevation pointing movements, and estimate the position of the Sun for each sun tracker coordinate. The analysis module uses the trajectory of the efficiency in the conversion of light energy into electricity, $[ts,\eta]$, and its associated movement trajectory, $[ts,\theta]$.

\begin{figure}[H]
	\centering
	\includegraphics[scale = 0.35]{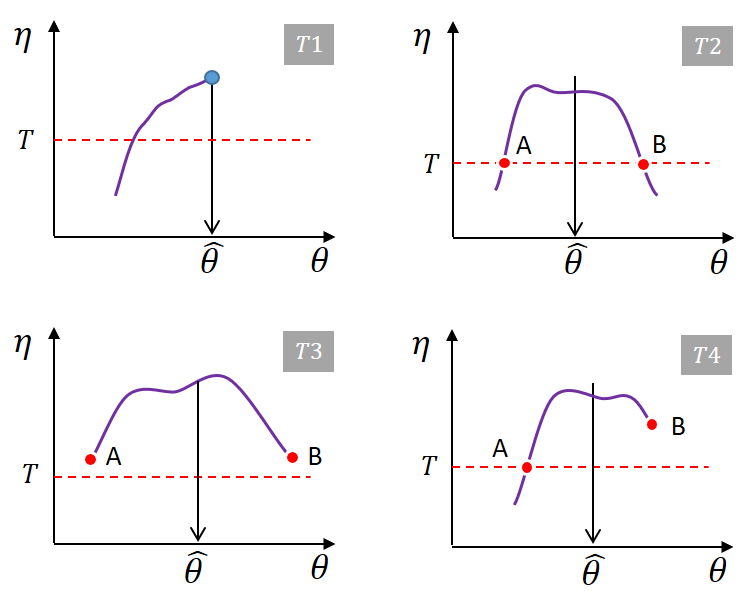}
	\caption{Types of efficiency trajectories.}
	\label{fig:TajectoryClasification}	
\end{figure}

Four types of expected trajectories are defined. They are based in the different shapes which can be obtained when slicing the power surface of the virtual cell. The analyzer module will try to classify the trajectory $[\eta,\theta]$ into one of the types from $T1$ to $T4$ by simple checks. Figure \ref{fig:TajectoryClasification} depicts the classes taken into account. If the trajectory does not correspond to any of the classes, it will be classified as \textit{rejected trajectory}. The threshold variable, $T$, is defined as a percentage of the maximum value of the efficiency ($90\%-95\%$). Notice that points A and B can be initial and final points or vice versa, depending of the sense of the movement and the path that the projection of the sun ray describes in the plane of the photovoltaic cell during the movement.

The types are described below:

\begin{itemize}
	
	\item Type T1: Off-center trajectory. This type of trajectory arises when the controller is in a start-up phase or after a cloudy period. The value of the sun tracker coordinate where the sun is located, $\hat{\theta}$, is chosen where the maximum efficiency is located.

	\item Type T2: Centered trajectory 1. This type of path is the expected when the controller is not in a start-up phase, but in normal operation. It is characterized by the following facts: initial and final points are below the threshold, $T$, and the point with maximum efficiency, is not A nor B. The points A and B are obtained by iterating from the beginning and ending points of the trajectory and checking when they exceed the threshold. The value of $\hat{\theta}$ is the mid-point between points A and B. The coordinate with maximum efficiency is not chosen because, although the trajectories are filtered, noise and perturbations can still deform the shape of the trajectories.

	\item Type T3: Centered trajectory 2. This type of path is the expected when the controller is not in a start-up phase, but in normal operation. Its characteristics are: initial and final points are over the threshold, $T$, and the point with maximum efficiency, is not A nor B (and is greater than the efficiency of A and B). The value of $\hat{\theta}$ is also chosen as the mid-point between points A and B.

	\item Type T4: Half-centered trajectory. This kind of trajectory will be obtained in a start-up phase of the controller at the beginning of the day. One of the extreme points is below the threshold and the other is over it. The point A is obtained by iterating from the point below the threshold and checking when it exceed the threshold. The value of  $\hat{\theta}$ is also chosen as the mid-point between points A and B.
	
\end{itemize}

There are situations during normal operation of the sun tracker in which the trajectories may not provide information. These cases are usually directly related with the start-up phase of the power inverter. It may take about two minutes for the power inverter to start the power injection once it begins to receive enough solar radiation, which means that at the beginning of the morning and during prolonged periods with clouds, the efficiency falls below a minimum threshold to work with. These low efficiency situations also must be taken into account by the classifier.

The resulting measured sun coordinates will not always be a good estimation because of the lack of information in the trajectories, but at least must point in the right direction. If this requirement is met, the control strategy will correct the discrepancies between Solar Equations and measurements over time, and the trajectories will have better information which will allow calculating better estimates.

\section {Control system } \label{sec:ControlSystem}

\begin{figure}[H]
	\centering
	\includegraphics[scale = 0.3]{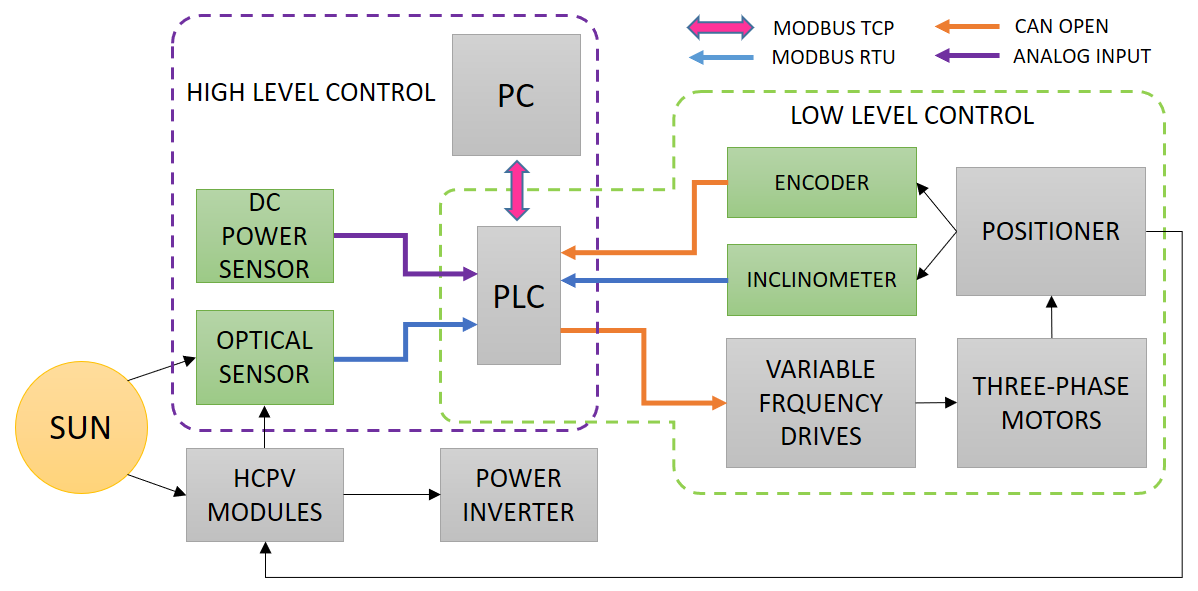}
	\caption{Scheme of the sun tracker control system.}
	\label{fig:ControlSystem}	
\end{figure}

The control system is divided in two layers. The low level layer deals with the control of the position of the two three-phase motors according to a given reference. This control is based on feedback of the motor shaft position by reading the pulses provided by the encoder and the angle provided by the inclinometer. The high level layer deals with the calculation of the set-points for the low level layer, that is, the values ​​that must be taken by the articular coordinates of the positioner to track the sun during the day. This is accomplished by feeding back the electric power produced by the solar tracker using the strategy described in Section \ref{sec:TrackingStrategy}. Figure \ref{fig:ControlSystem} shows a scheme of the sun tracker control system with all the devices and their relationships.

The measures of electric power are provided by a custom DC power meter that generates a low voltage signal proportional to the electric power. It is connected to an analog input of the PLC. There is also an optical sensor mounted on the wing of the sun tracker that provides measures of solar irradiance (DNI). It communicates with the PLC over a Modbus RTU bus. This last measurement is necessary to compute the efficiency, although the tracking strategy would also work directly with the power readings.

As can be seen in Fig. \ref{fig:ControlSystem}, the PLC is shared by the two control layers. The PLC takes care of all the tasks regarding sensors reading and values conversion to engineering units as well as low level control. The PLC cycle time is less than 10 ms, and the highest sensor period reading is 125 ms (optical sensor). The high level layer comprises the PLC and a PC, which is in charge of perform time and memory consuming tasks. These tasks that are too demanding to be ran in the PLC are trajectories storage, non-causal FIR filtering, analysis, and prediction using Solar Equations, etc. 

The PC executes an application that performs functions of monitoring measurements and other variables, supervision and data logging on disk. Some parameters of the program that the PLC executes can be changed using this application. These configuration parameters are related to motion (motor speeds, resting position, sun tracker coordinates motion limits, etc.), and to the system (PLC clock setting, resting schedule, set encoder zero, etc.). It also allows to operate the sun tracker in jog mode and to switch between several control strategies (closed- loop with power feedback, open- loop with Solar Equations, etc.). The information exchange between PC and PLC is done over a Modbus TCP field-bus with a period of 250 ms.

\section{Experimental results} \label{sec:Results}

Some considerations have to be taken into account for the presented tests. The zero of the orientation encoder of the sun tracker was adjusted in a coarse manner using visual references. No precise calibration was performed in order to test the validity of the proposed control algorithm. In all the experiments the value of $\beta$ is equal to 0.5 degrees. In order to avoid collisions of the wing with the devices attached to the pole of the sun tracker, the controller has a software limit in the elevation coordinate, $\theta_{ele}$, equal to 20 degrees. Therefore, until the Sun elevation reaches 20 degrees the sun tracker is stuck at this elevation and can not perform elevation movements. The main source of assembly uncertainty in the sun tracker is caused by a not too precise mounting of the solar modules onto the wing. Their catching surfaces do not lie in a common plane, but there are deviations up to half a degree in different orientations. This causes electrical mismatches which decrease the final efficiency of the sun tracker, and also can cause efficiency variations depending on where the concentrated sun beam projects onto the virtual cell (as it will be a different point on each module).

\subsection{SUNNY DAY}

\begin{figure}[H]
	\centering
	\includegraphics[scale = 0.55]{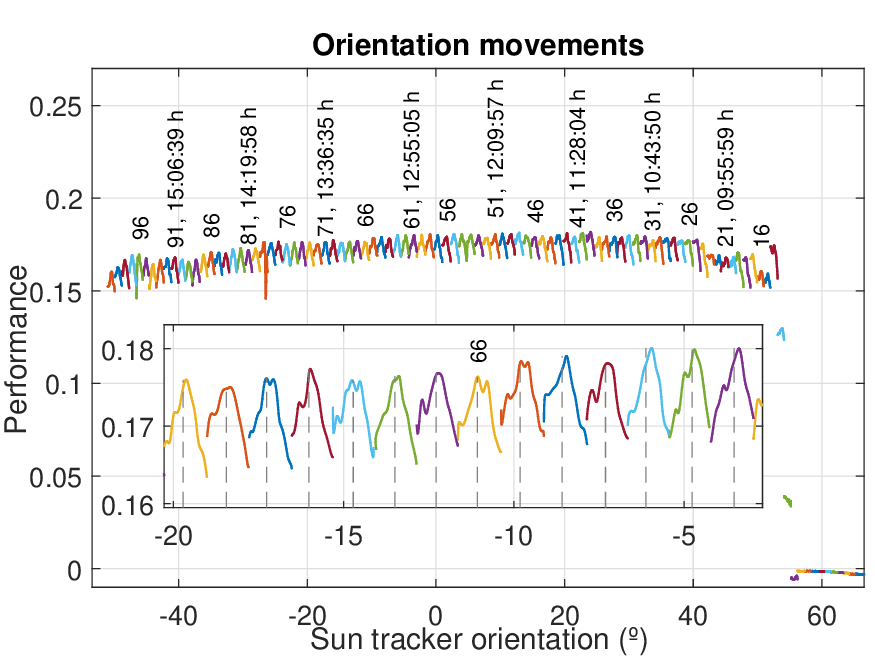}
	\caption{Efficiency trajectories associated to orientation movements. Each movement is represented with a different color for the sake of clarity. In the detail view the vertical dashed lines represent the estimated sun position in sun tracker elevation coordinate.}
	\label{fig:MvtoORI_1}	
\end{figure}

\begin{figure}[H]
	\centering
	\includegraphics[scale = 0.55]{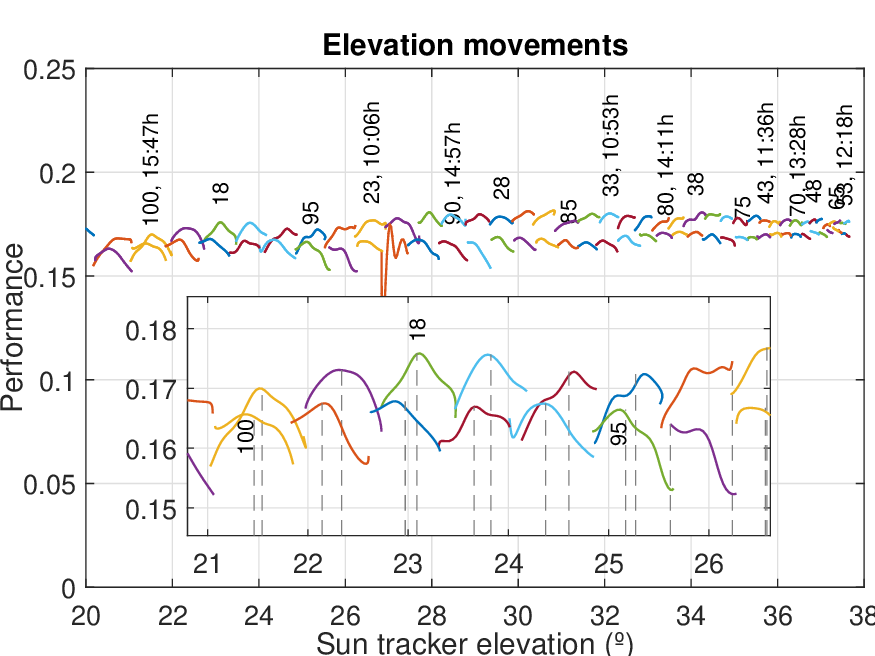}
	\caption{Efficiency trajectories associated to elevation movements. Each movement is represented with a different color for the sake of clarity. In the detail view the vertical dashed lines represent the estimated sun position in sun tracker elevation coordinate.}
	\label{fig:MvtoELE_1}	
\end{figure}

The test was carried out on 2/10/2020 from 8 h to 16 h (solar time). The efficiency trajectories sampled during the orientation and elevation movements are shown in Figs. \ref{fig:MvtoORI_1} and \ref{fig:MvtoELE_1} respectively. The text over the trajectories represents the corresponding trajectory number (starting at one) associated to a sun tracker movement, and the time instant at it was performed (solar time). Each movement is represented with a different color for the sake of clarity. It can be seen how the first orientation movements in the morning have null efficiency trajectories due to collision avoiding, and also that there are no elevation movements. In the detail view of Figs. \ref{fig:MvtoORI_1} and \ref{fig:MvtoELE_1}, the vertical dashed lines represent the estimated sun position during each scan movement in sun tracker orientation coordinate and elevation coordinate, respectively.

\begin{figure}[H]
	\centering
	\includegraphics[scale = 0.55]{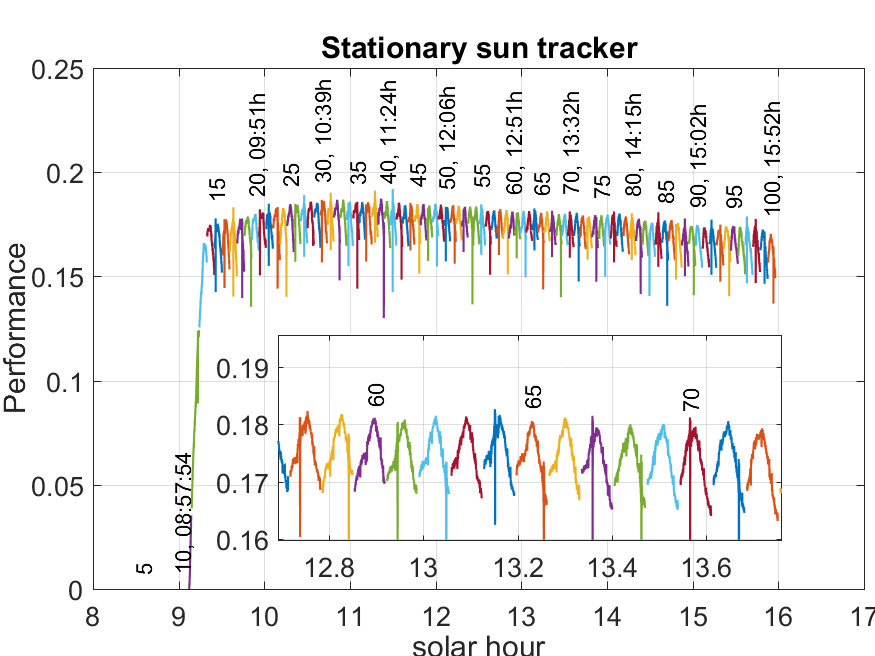}
	\caption{Efficiency in the time interval between pointing movements.}
	\label{fig:MvtoSOL_1}
\end{figure}

Figure \ref{fig:MvtoSOL_1} shows the sampled efficiency trajectories while the sun tracker is motionless in the time interval between pointing movements. It can be seen how the trajectories in the time interval from 9 h to 10 h, once the power inverter starts the power injection, are very unbalanced (not symmetrical) and start to become centered after some movements, acquiring the expected shape.

\begin{figure}[H]
\centering
\includegraphics[scale = 0.55]{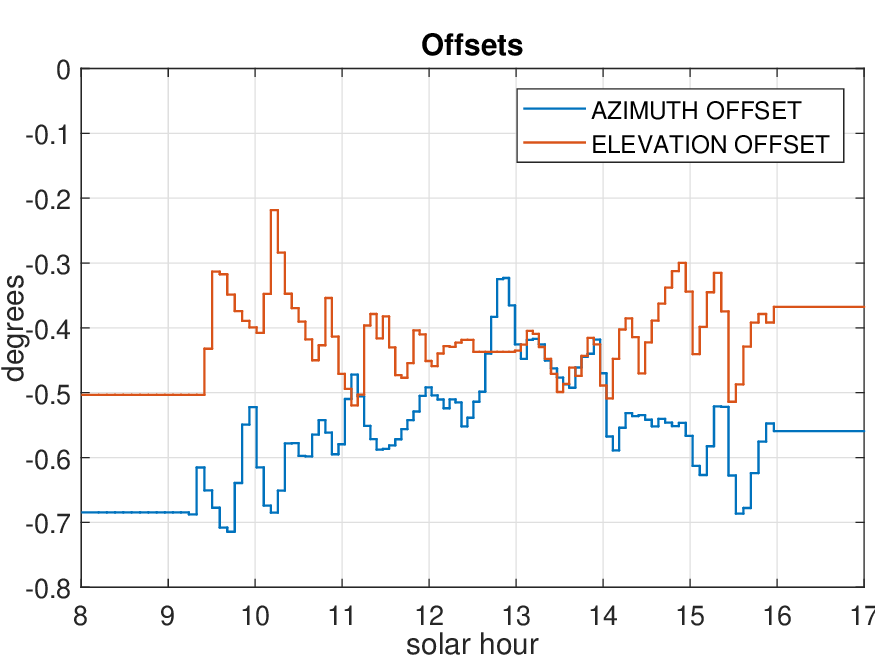}
\caption{Calculated azimuth and elevation offsets used to correct the Solar Equations.}
\label{fig:OFFSETS_1}
\end{figure}

\begin{figure}[H]
	\centering
	\includegraphics[scale = 0.55]{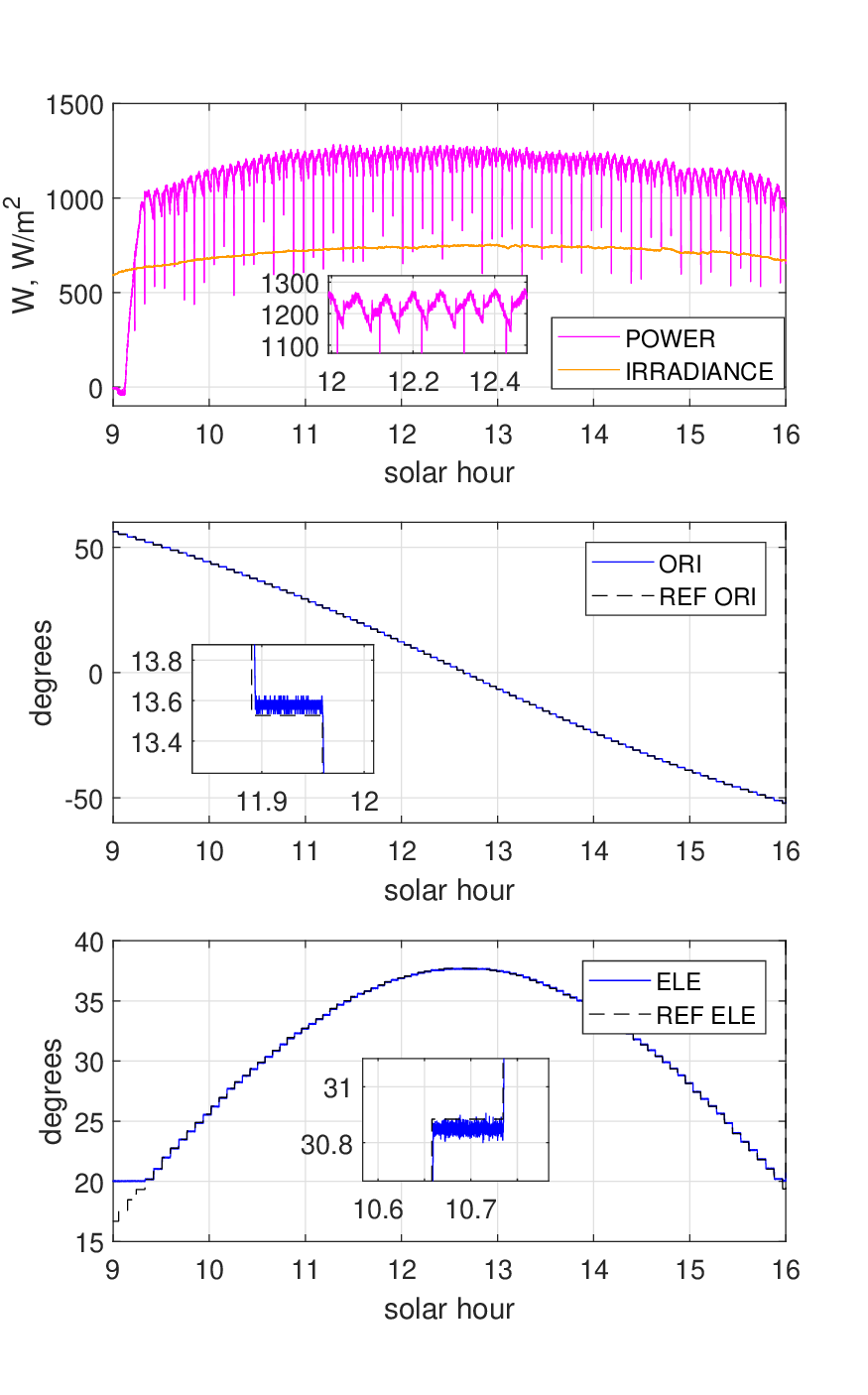}
	\caption{Produced DC power and sun tracker trajectories in its own coordinate system for a sunny day. The detail views show the accuracy of the aiming system to reach the references.}
	\label{fig:POTENCIA_1}	
\end{figure}

The computed Azimuth and elevation offsets which are used to correct the Solar Equations are depicted in Fig. \ref{fig:OFFSETS_1}. The initial values for these offsets are the last values stored the previous day. Figure \ref{fig:POTENCIA_1} shows the electric power produced in the DC side of the power inverter along with the orientation and elevation trajectories the sun tracker follows, in sun tracker coordinates. The periodic power drops are caused by the maximum power point tracking (MPPT) algorithm of the power inverter.

\subsection{Cloudy day}

\begin{figure}[H]
	\centering
 	\includegraphics[scale = 0.55]{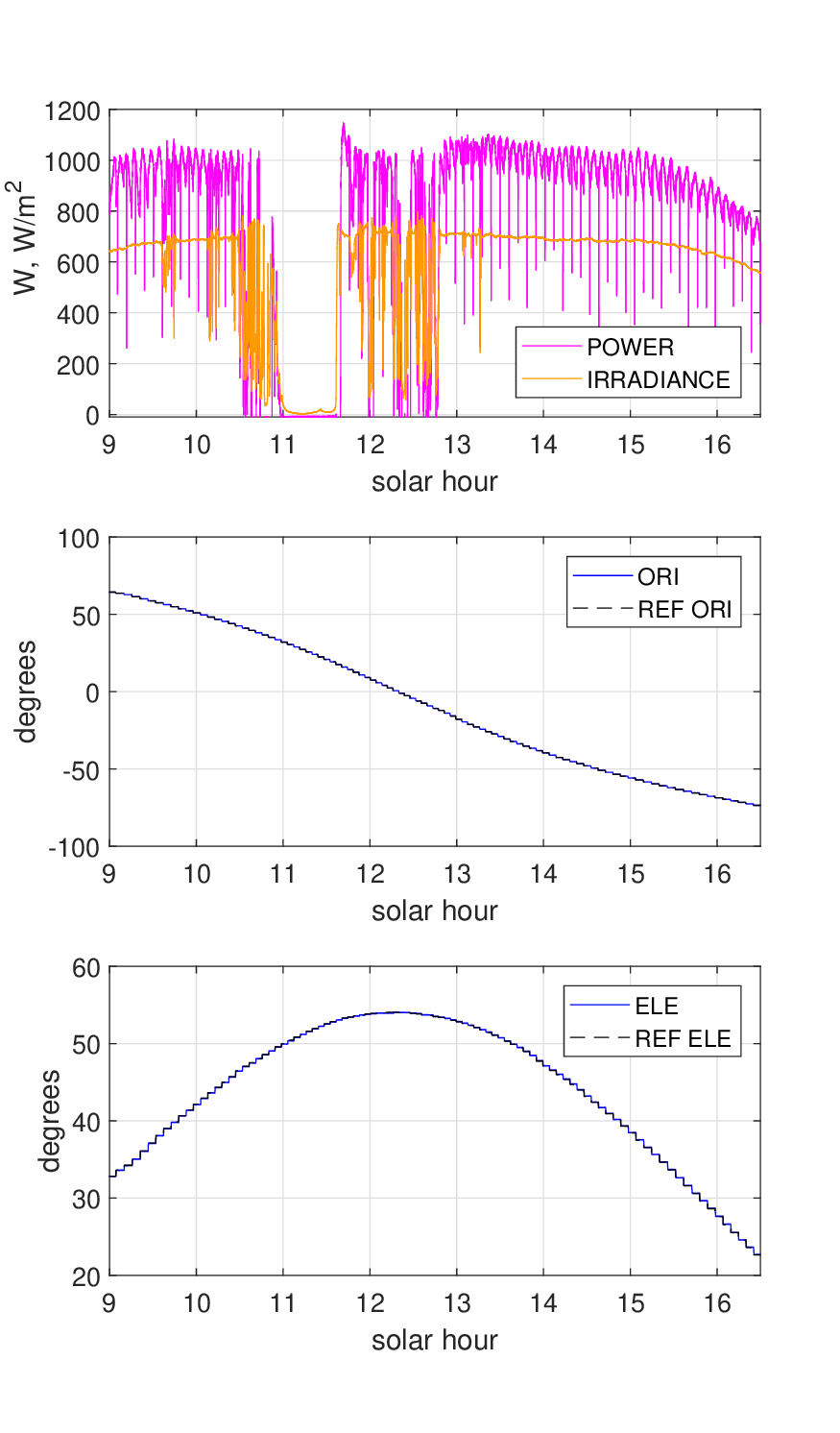}
	\caption{Produced DC power and sun tracker trajectories in its own coordinate system for a cloudy day.}
	\label{fig:POTENCIA_2}	
\end{figure}

\begin{figure}[H]
	\centering
	\includegraphics[scale = 0.55]{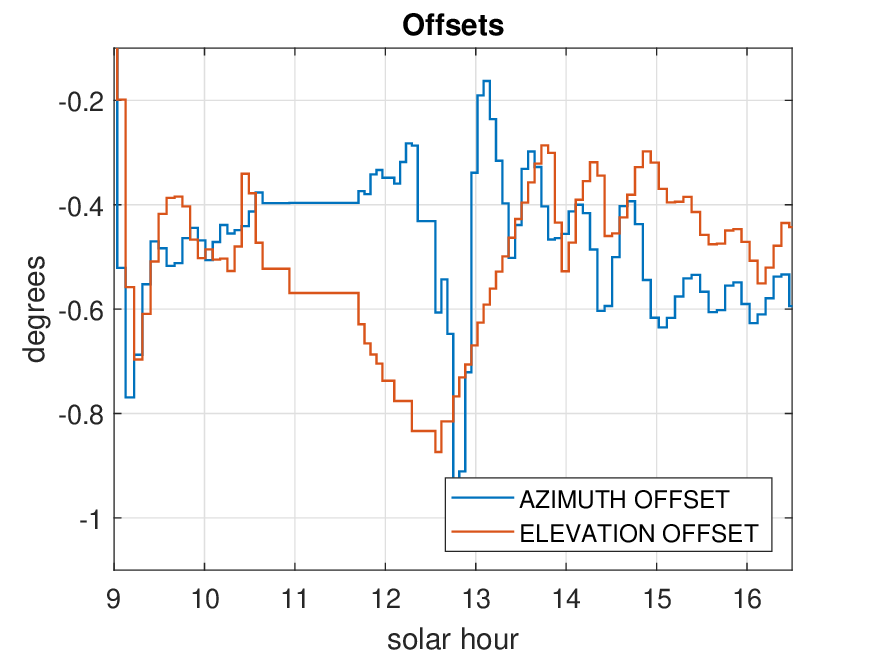}
	\caption{Calculated azimuth and elevation offsets for a cloudy day.}
	\label{fig:OFFSETS_2}	
\end{figure}

Figures \ref{fig:POTENCIA_2} and \ref{fig:OFFSETS_2} show the sun tracking results of an experiment carried out on 09/17/2019, a day with a relatively long cloudy period. It can be observed in Fig. \ref{fig:POTENCIA_2} that between 11 h and 11:45 h the irradiance becomes zero due to the passing of a cloud, while Fig. \ref{fig:OFFSETS_2} shows how during this cloudy time period the PI controllers are not actualized (when irradiance is zero, the efficiency is also set to zero and the trajectories are classified as low efficiency). Therefore, the last computed azimuth and elevation offsets are kept and continues correcting the Solar Equations during the passing of the cloud. This means that the sun tracker keeps moving as usual during the cloudy period, but its tracking accuracy will be reducing with the time.  When the cloudy time period ends, the next sampled trajectories are valid again (they are not classified as low efficiency) and the controller continues computing new offsets.

\subsection{Behaviour with a higher degree of uncalibration} \label{sec:highDegreeUncalibration}

\begin{figure}[H]
	\centering
	\includegraphics[scale = 0.6]{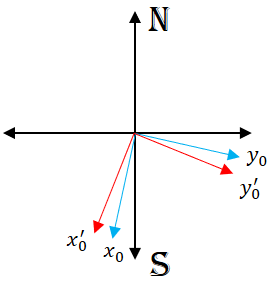}
	\caption{Reference systems. The Azimuth angle provided by the solar Equations is referred to geographical North. The sun tracker platform reference frame, $\{0\}$, is rotated in order to decrease the accuracy when pointing to geographical south.}
	\label{fig:ReferenceSystem}	
\end{figure}

With the objective of testing the proposed control strategy under more unfavorable conditions, a small change is introduced in the rotation matrix that relates the Solar Equations frame with the sun tracker platform frame. The reference system $\{0\}$ is the frame associated to the platform of the sun tracker. The transformation between both reference systems consist of a rotation around the Z-axis equal to 180 degrees in order to point to geographical south. However the frame $\{0\}$ will not align exactly to the south due to inaccuracies in the installation of the platform, as exemplified in Fig. \ref{fig:ReferenceSystem}. For this test, the reference system $\{0\}$ will be rotated around the Z-axis in order to decrease the accuracy when pointing to geographical south, so that it becomes the reference system $\{0'\}$. This rotation matrix is a parameter of the controller, and does not mean that the actual sun tracker is oriented exactly to this direction. 

\begin{figure}[H]
	\centering
	\includegraphics[scale = 0.55]{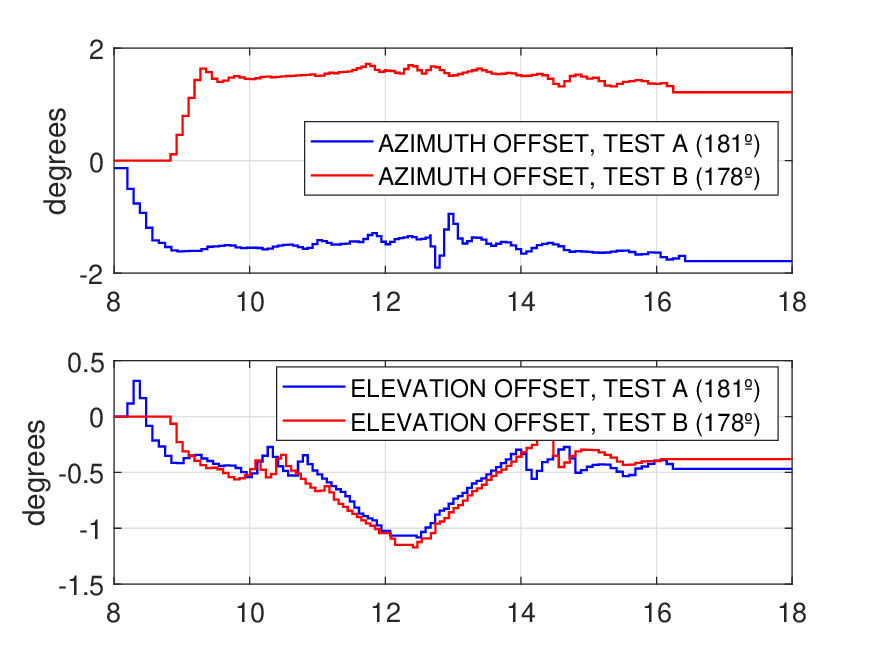}
	\caption{Calculated offsets for tests A and B.}
	\label{fig:Comparacion3}	
\end{figure}

\begin{figure}[H]
	\centering
	\includegraphics[scale = 0.55]{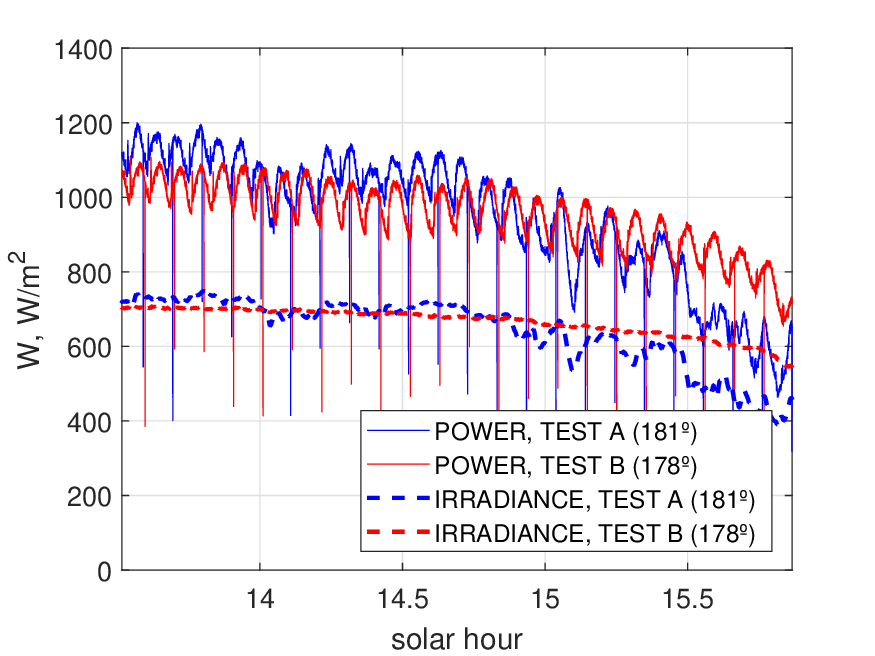}
	\caption{Power produced in the tests A and B.}
	\label{fig:Comparacion2}	
\end{figure}

Two experiments are carried out. The first one was carried out on 10/03/2019 and the rotation about the Z-axis is changed to 181 degrees (Test A). The second one was carried out on 10/09/2019 and the rotation is changed to 178 degrees (Test B). The resulting azimuth and elevation offsets are shown in Fig. \ref{fig:Comparacion3}. In the 178 degrees case, the azimuth offset, initially set to zero for both tests, converges smoothly to a positive value while in the 181 degrees case it converges to a negative value, since the first rotation is less than 180 degrees and the second is greater. Figure \ref{fig:Comparacion2} depicts the DC power produced during both tests. It can be seen that there is no important difference of efficiency between the tests, as the control strategy is able to adapt to the introduced inaccuracies in the installation of the platform.

\subsection{Comparison with an open-loop control strategy}

\begin{figure}[H]
	\centering
	\includegraphics[scale = 0.6]{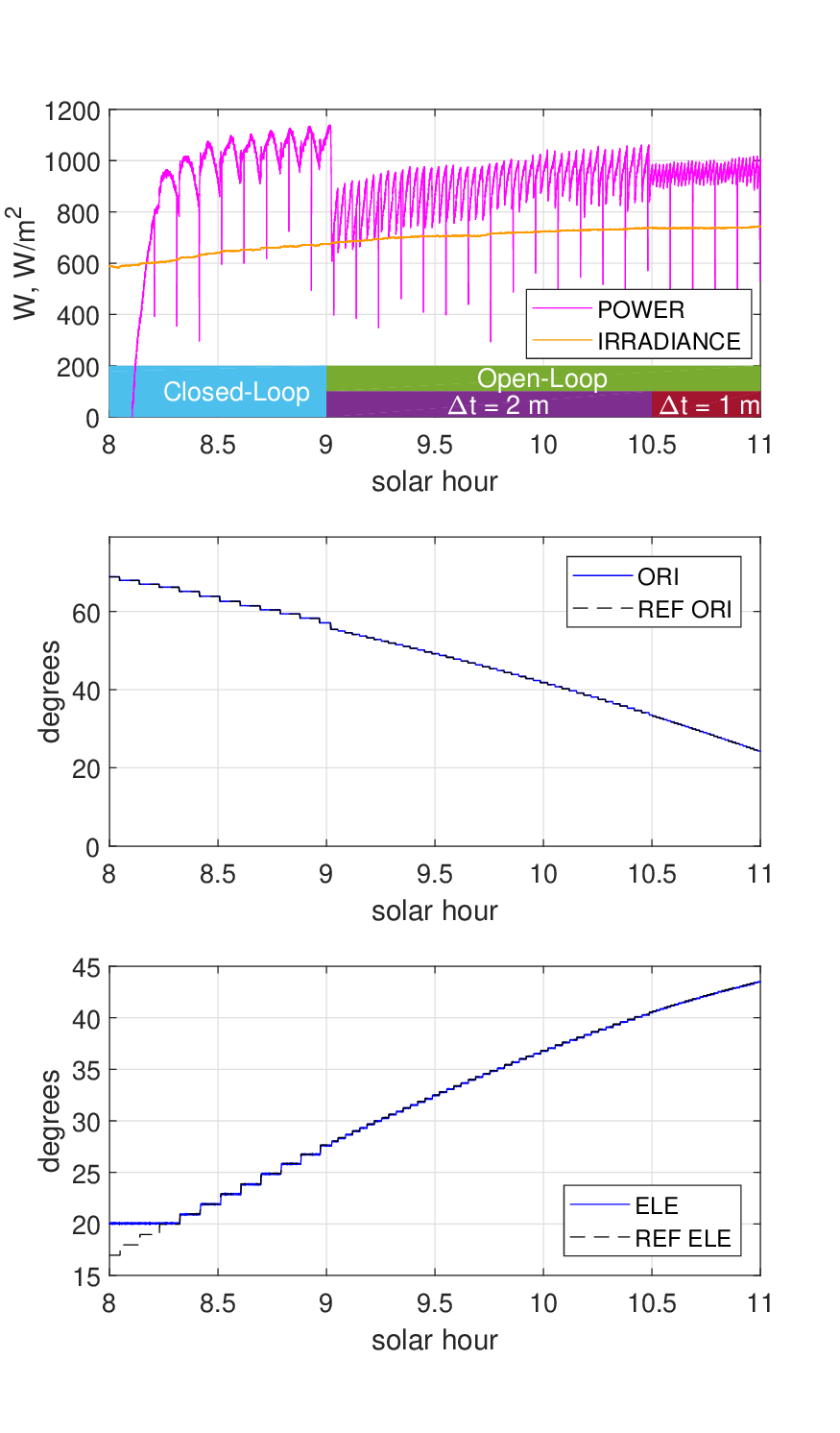}
	\caption{Comparison between the power feedback closed-loop and the open-loop control schemes.}
	\label{fig:Comparacion}	
\end{figure}

A comparison between the closed-loop with power feedback control strategy and an open-loop by means of Solar Equations strategy is shown in Fig. \ref{fig:Comparacion}. The date of the test is 10/08/2019. In both cases, the sun tracker has been uncalibrated on purpose in order to test the efficiency of the control strategies under conditions that simulates an inaccurate installation of the platform. The un-calibration consist on setting the rotation around the Z-axis of the reference frame of the platform to a value of 181 degrees, in the same way it has been explained in section \ref{sec:highDegreeUncalibration}. For the open-loop case, the time between pointing movements has been set to 1 min and 2 min in order to see how the produced DC power is affected in conjunction with the rotation of the Z-axis of the platform reference frame.

It can be observed how in the case of closed-loop strategy with power feedback, the generated DC power is above the open-loop strategy case even with less received irradiance. The measured mean efficiency in closed-loop is approximately equal to 0.17, while in open-loop is approximately equal to 0.135 when moving every 2 min and 0.14 when moving every 1 min. This translates into a efficiency increase of approximately 21\% when using the proposed closed-loop control strategy on a non-calibrated sun tracker.

\section {Conclusions} \label{sec:Conclusiones}

A closed-loop control strategy which is able to correct the different sources of uncertainty that can be found in a sun tracker in order to increase pointing accuracy, and therefore the generated electric power, without the need of costly installation or calibration has been developed. This is an advantage over classic open-loop and closed-loop tracking strategies, which need to be very well set-up in order to have an optimal efficiency.

It has been proven with experimental tests that proposed tracking strategy works for relatively low levels of misalignment of the platform reference system with respect geographical south. If the degree of misalignment were higher it would be necessary to resort to techniques to locate the Sun and calculate a initial values for the Azimuth and elevation offsets before executing the control algorithm.

It has been found an unexpected ripple in the generated electric power. By looking the power surface obtained experimentally with a constant load (see Fig. \ref{fig:PowerSurface}) one would think that, using a $\beta$ value of 0.5 degrees, the generated squared region in the plane of the virtual cell would be a constant maximum efficiency region, but it is not the case. When the constant load is replaced by the power inverter, the power surface stretches and therefore the acceptance angle contracts. The ripple could be decreased by giving the parameter $\beta$ smaller values, but in turn more orientation and elevation movements would be performed throughout the day. Regardless of the ripple, the proposed tracking strategy has proved to have higher mean generated electric power than a pure open-loop tracking strategy when the system is not well calibrated.

\section*{Acknowledgment}

This work was supported by the Spanish Ministry of Economy and Competitiveness [grant DPI2016-79444-R]. This support is gratefully acknowledged. 

\bibliography{mybibfile_mod_red}

\end{document}